\def\~{{$\tilde{\phantom{a}}$}}

\documentclass [12pt] {article}
\usepackage{epsfig}
\usepackage{color}

\def\thebibliography#1{\section{References}\markboth
 {REFERENCES}{REFERENCES}\list
 {[\arabic{enumi}]}{\settowidth\labelwidth{[#1]}\leftmargin\labelwidth
 \advance\leftmargin\labelsep
 \usecounter{enumi}}
 \def\newblock{\hskip .11em plus .33em minus -.07em}
 \sloppy
 \sfcode`\.=1000\relax}
\def\upcite#1{\raise6pt\hbox{\scriptsize
\cite{#1}}}
  \def\lsim{\mathrel {\vcenter {\baselineskip 0pt \kern 0pt
    \hbox{$<$} \kern 0pt \hbox{$\sim$} }}}
    \def\gsim{\mathrel {\vcenter {\baselineskip 0pt \kern 0pt
    \hbox{$>$} \kern 0pt \hbox{$\sim$} }}}

\def\hline{\noalign{\hrule \vskip2pt}}

\def\|{\ifmmode\Vert\else \char`\|\fi}
\ifx\oldzeta\undefined                          
  \let\oldzeta=\zeta                            
  \def\zzeta{{\raise 2pt\hbox{$\oldzeta$}}}     
  \let\zeta=\zzeta                              
\fi

\ifx\oldchi\undefined                           
  \let\oldchi=\chi                              
  \def\cchi{{\raise 2pt\hbox{$\oldchi$}}}       
  \let\chi=\cchi                                
\fi

\def\frac#1#2{{#1 \over #2}}

\def\half{\ifinner {\scriptstyle {1 \over 2}}
   \else {1 \over 2} \fi}

\def\simge{\mathrel{%
   \rlap{\raise 0.511ex \hbox{$>$}}{\lower 0.511ex \hbox{$\sim$}}}}
\def\simle{\mathrel{
   \rlap{\raise 0.511ex \hbox{$<$}}{\lower 0.511ex \hbox{$\sim$}}}}

\def\buildchar#1#2#3{{\null\!                   
   \mathop#1\limits^{#2}_{#3}                   
   \!\null}}                                    
\def\overcirc#1{\buildchar{#1}{\circ}{}}

\def\slashchar#1{\setbox0=\hbox{$#1$}           
   \dimen0=\wd0                                 
   \setbox1=\hbox{/} \dimen1=\wd1               
   \ifdim\dimen0>\dimen1                        
      \rlap{\hbox to \dimen0{\hfil/\hfil}}      
      #1                                        
   \else                                        
      \rlap{\hbox to \dimen1{\hfil$#1$\hfil}}   
      /                                         
   \fi}                                         %

\def\subrightarrow#1{
  \setbox0=\hbox{
    $\displaystyle\mathop{}
    \limits_{#1}$}
  \dimen0=\wd0
  \advance \dimen0 by .5em
  \mathrel{
    \mathop{\hbox to \dimen0{\rightarrowfill}}
       \limits_{#1}}}                           

\def\overlay#1#2{\ifmmode%
\setbox0=\hbox{$#1$}%
\setbox1=\hbox to\wd0{\hss$#2$\hss}\else%
\setbox0=\hbox{#1}%
\setbox1=\hbox to\wd0{\hss#2\hss}\fi%
#1\hskip-\wd0\box1 }

\def\pmb#1{\leavevmode\setbox0=\hbox{#1}%
\kern-.02em\copy0\kern-\wd0
\kern.04em\copy0\kern-\wd0
\kern-.02em\raise.04em\box0 }

\def\vereq#1#2{\lower3pt\vbox{\baselineskip1.5pt \lineskip1.5pt
\ialign{$\m@th#1\hfill##\hfil$\crcr#2\crcr\sim\crcr}}}

\def\tensor#1{\protect\@ontopof{#1}{\leftrightarrow}{1.15}\mathord{\box2}}
\def\overstar#1{\protect\@ontopof{#1}{\ast}{1.15}\mathord{\box2}}
\def\overdots#1{\protect\@ontopof{#1}{\cdots}{1.0}\mathord{\box2}}
\def\overcirc#1{\protect\@ontopof{#1}{\circ}{1.2}\mathord{\box2}}
\def\loarrow#1{\protect\@ontopof{#1}{\leftarrow}{1.15}\mathord{\box2}}
\def\roarrow#1{\protect\@ontopof{#1}{\rightarrow}{1.15}\mathord{\box2}}

\def\@ontopof#1#2#3{%
{\mathchoice
{\@@ontopof{#1}{#2}{#3}\displaystyle\scriptstyle}%
{\@@ontopof{#1}{#2}{#3}\textstyle\scriptstyle}%
{\@@ontopof{#1}{#2}{#3}\scriptstyle\scriptscriptstyle}%
{\@@ontopof{#1}{#2}{#3}\scriptscriptstyle\scriptscriptstyle}%
}%
}

\def\@@ontopof#1#2#3#4#5{%
\setbox0=\hbox{$#4#1$}%
\setbox1=\hbox{$#5#2$}%
\setbox2=\hbox{}\ht2=\ht0 \dp2=\dp0 %
\ifdim\wd0>\wd1 %
\setbox1=\hbox to\wd0{\hss\box1\hss}%
\mathord{\rlap{\raise#3\ht0\box1}\box0}%
\else   %
\setbox1=\hbox to.9\wd1{\hss\box1\hss}%
\setbox0=\hbox to\wd1{\hss$#4\relax#1$\hss}%
\mathord{\rlap{\copy0}\raise#3\ht0\box1}%
\fi
}%

\def\lambdabar{\protect\@lambdabar}
\def\@lambdabar{%
\relax
\bgroup
\def\@tempa{\hbox{\raise.73\ht0
\hbox to0pt{\kern.25\wd0\vrule width.5\wd0
height.1pt depth.1pt\hss}\box0}}%
\mathchoice{\setbox0\hbox{$\displaystyle\lambda$}\@tempa}%
{\setbox0\hbox{$\textstyle\lambda$}\@tempa}%
{\setbox0\hbox{$\scriptstyle\lambda$}\@tempa}%
{\setbox0\hbox{$\scriptscriptstyle\lambda$}\@tempa}%
\egroup
}

\def\corresponds{{\lower.2ex\hbox{=}}{\rm\kern-.75em^\triangle}}
\def\succsim{\succ\kern-.9em_\sim\kern.3em}
\def\precsim{\prec\kern-1em_\sim\kern.3em}
\def\slantfrac#1#2{\kern1em^{#1}\kern-.3em/\kern-.1em_{#2}}

\begin{document}

\begin{center}
{\Large\bf Energy Balance in an Electrostatic Accelerator}
\\

\medskip

Max S.~Zolotorev
\\
{\sl Center for Beam Physics, Lawrence Berkeley National Laboratory,
Berkeley, CA 94720}
\\
Kirk T.~McDonald
\\
{\sl Joseph Henry Laboratories, Princeton University, Princeton, NJ 08544}
\\
(Feb.~1, 1998)
\end{center}

\section{Problem}

The principle of an electrostatic accelerator is that when
a charge $e$ escapes from a conducting plane that supports a uniform
electric field of strength $E_0$, then the charge gains energy $eE_0d$ as it
moves distance $d$ from the plane.  Where does this energy come from?

Show that the mechanical energy gain of the electron is balanced by the decrease
in the electrostatic field energy of the system.


\section{Solution}

Once the charge has reached distance $d$ from the plane, the static
electric field ${\bf E}_e$ at an arbitrary point {\bf r} due to the charge
can be calculated by summing the field of the charge plus its image
charge,
\begin{equation}
{\bf E}_e({\bf r},d) = {e{\bf r}_1 \over r_1^3} - {e{\bf r}_2 \over r_2^3},
\label{eq1}
\end{equation}
where ${\bf r}_1$ (${\bf r}_2$) points from the charge (image) to the
observation point {\bf r}, as illustrated in Fig.~1.
The total electric field is then $E_0 \hat{\bf z} + {\bf E}_e$.

\begin{figure}[htp]  
\begin{center}
\includegraphics[width=3in, angle=0, clip]{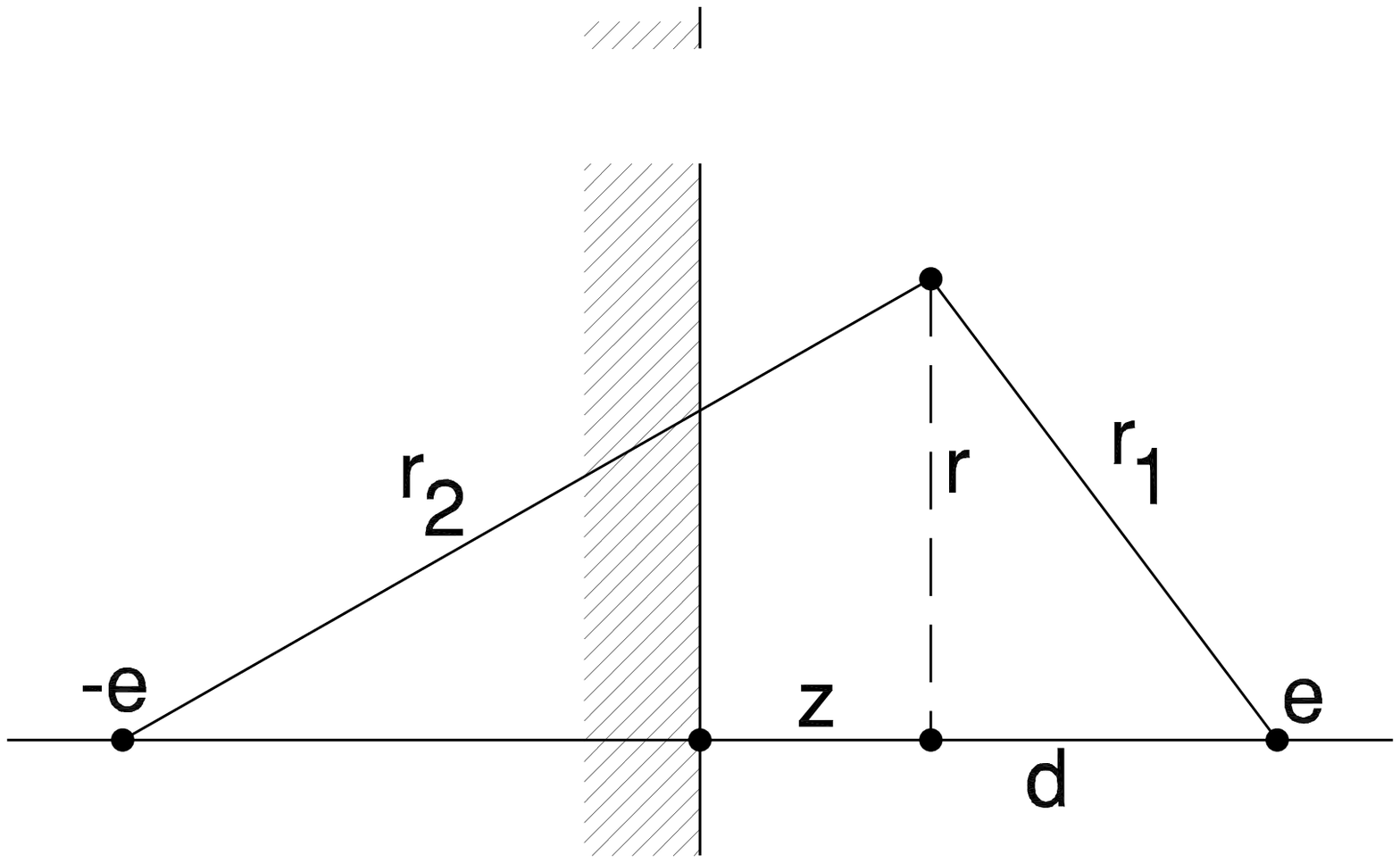}
\parbox{5.5in} 
{\caption[ Short caption for table of contents ]
{\label{fig1} The charge $e$ and its image charge $-e$ at positions
$(r,\theta,z) = (0,0,\pm d)$ with respect to a conducting plane at $z = 0$.
Vectors ${\bf r}_1$ and ${\bf r}_2$ are directed from the charges to the
observation point $(r,0,z)$.
}}
\end{center}
\end{figure}

It turns out to be convenient to use a cylindrical coordinate system, where
the observation point is ${\bf r} = (r,\theta,z) = (r,0,z)$, 
and the charge is at $(0,0,d)$. Then,
\begin{equation}
r_{1,2}^2 = r^2 + (z \mp d)^2.
\label{eq2}
\end{equation}

The part of the electrostatic field energy that varies with the position of
the charge is the interaction term (in Gaussian units),
\begin{eqnarray}
U_{\rm int} & = & \int {E_0 \hat{\bf z} \cdot {\bf E_e} \over 4\pi }d{\rm Vol}
\nonumber \\
& = & {e E_0 \over 4\pi} \int_0^\infty dz  \int_0^\infty \pi dr^2
\left( {z - d \over [r^2 + (z - d)^2]^{3/2} } - 
        {z + d \over [r^2 + (z + d)^2]^{3/2} } \right)
\nonumber \\
& = & {e E_0 \over 4} \int_0^\infty dz \left( \left\{ \begin{array}{ll}
    2 & \mbox{if}\ z > d \\
   -2 & \mbox{if}\ z < d 
\end{array} \right\} - 2 \right)
\nonumber \\
& = & - e E_0 \int_0^d dz = -e E_0 d.
\label{eq3}
\end{eqnarray}
When the particle has traversed a potential difference $V = E_0 d$, it has 
gained energy $eV$ and the electromagnetic field has lost the same energy.

In a practical ``electrostatic'' accelerator, the particle
is freed from an electrode at potential $-V$
and emerges with energy $eV$ in a region of zero potential.  However, the
particle could not be moved to the negative electrode from a region of zero
potential by purely electrostatic forces unless the particle lost energy
$eV$ in the process, leading to zero overall energy change.  
An ``electrostatic'' accelerator must have an essential component
(such as a battery) 
that provides a nonelectrostatic force that can absorb the energy extracted 
from the electrostatic field while moving the charge from
potential zero, so as to put the charge at rest at potential $-V$ prior to
acceleration.

\end{document}